\documentclass[11pt,english,aps]{revtex4-1}
\usepackage[T1]{fontenc}
\usepackage[latin9]{inputenc}
\usepackage{amsmath}
\usepackage{amssymb}
\usepackage{babel}
\begin{document}

\title{Constructive Cavity Method}

\author{Simone Franchini}

\address{Sapienza Universita di Roma, Piazza Aldo Moro 1, 00185 Roma, Italy
\\
CNR-ISTC, Via Gian Domenico Romagnosi 18, 00196 Roma, Italy}

\email{corresponding author: simone.franchini@yahoo.it}

\begin{abstract}

We show that the functional appearing in the celebrated Parisi formula
for the free energy of the Sherrington-Kirkpatrick model can be found from the incremental free energy obtained by Cavity Method if one assumes that
the state is a product of independent Random Energy models. 
\end{abstract}
\maketitle

A trial state for the Sherrington-Kirkpatrick (SK) model \cite{PMM,Charbonneau,Franchini,Franchini2021,FranchiniSPA2,FranchiniUniversal,ASS,Bolt,GuerraGG,PakUltra,PaK} is shown where the spin group is partitioned
into a number of sub-groups that behave independently like Random Energy Models (REM, see in \cite{Kurkova1,Arous-Kupsov}). 
In particular, we show that the probability measure associated to $L$ levels of Replica Symmetry Breakings (RSB, see in \cite{PMM,Charbonneau}) is equivalent (in distribution) to the product of $L$ independent REM Gibbs measures. The ansatz is tested by computing the corresponding functional from the expression of the incremental free energy that one obtains from Cavity Method \cite{PMM,ASS,Bolt,PaK}, and it indeed provides the correct Parisi functional.
The present letter is also intended to show the minimal path to obtain the Parisi functional from the Cavity Method and REM Universality \cite{FranchiniSPA2, Arous-Kupsov} by reducing to essential the Kernel Theory of RSB presented in \cite{Franchini,Franchini2021,FranchiniSPA2,FranchiniUniversal}. 

\section{Lower bound via cavity method}

We start by introducing the basic notation and recall the free energy
functional obtained from Cavity method. Let consider a finite spin
system of $N$ spins, governed by the SK Hamiltonian 
\begin{equation}
H\left(\sigma\right):=\frac{1}{\sqrt{N}}\sum_{i<j}\sigma_{i}J_{ij}\sigma_{j}.
\end{equation}
with $J$ instance of a Gaussian random matrix with normal independent
entries of unitary variance. As usual, we define the partition function 
and the associated Gibbs measure
\begin{equation}
Z:=\sum_{\sigma}\exp\left(-\beta H\left(\sigma\right)\right)
\ \ \ 
\mu\left(\sigma\right):=\frac{1}{Z}\exp\left(-\beta H\left(\sigma\right)\right)
\end{equation}
The free energy density is written in term of the pressure 
\begin{equation}
p:=\lim_{N\rightarrow\infty}\frac{1}{N}\log Z.
\end{equation}
Following the approach of Aizenmann, Sims and Starr (see for example in \cite{ASS,Bolt,PaK})
we define the cavity variables, i.e., the cavity field and the correction term:
\begin{equation}
\tilde{x}\left(\sigma\right):=\frac{1}{\sqrt{N}}\sum_{i}\tilde{J}_{ii}\sigma_{i} 
\ \ \ \
\tilde{y}\left(\sigma\right):=\frac{1}{N}\sum_{i<j}\sigma_{i}\tilde{J}_{ij}\sigma_{j}=\frac{1}{\sqrt{N}}\tilde{H}\left(\sigma\right)
\end{equation}
The correction term is proportional to the Hamiltonian, but with a new independent noise matrix $\tilde{J}$. Apart from vanishing terms in the large $N$ limit the incremental pressure is 
\begin{equation}
A\left(\mu\right):=\log2+\log\,\sum_{\sigma}\mu\left(\sigma\right)\cosh\left(\beta\tilde{x}\left(\sigma\right)\right)-\log\,\sum_{\sigma}\mu\left(\sigma\right)\exp\left(\beta\tilde{y}\left(\sigma\right)\right)
\end{equation}
and can be shown \cite{ASS,Bolt,PaK} that the actual pressure satisfies
\begin{equation}
p\geq\liminf_{N\rightarrow\infty}\,A\left(\mu\right)
\end{equation}
giving a lower bound that is proven exact for the SK model in the
thermodynamic limit. 

\section{Index notation}

In the previous we dropped the explicit dependence
from the noise matrices, also, for notation convenience we introduce
a labeling for the support of $\mu$. In our case it is the binary
spin kernel
\begin{equation}
\Omega^{N}:=\left\{ \tau^{\alpha}:\,1\leq\alpha\leq2^{N}\right\} 
\ \ \ 
\tau^{\alpha}:=\left\{ \tau_{i}^{\alpha}\in\Omega:\,1\leq i\leq N\right\} .
\end{equation}
and $\Omega:=\{-1,1\}$. Then, the measure is written as a mixture of atomic product measures
\begin{equation}
\mu\left(\sigma\right)=\sum_{\alpha}\mu^{\alpha}\prod_{i}I\left(\sigma_{i}=\tau_{i}^{\alpha}\right)
\end{equation}
where the points $\mu^{\alpha}\in\left[0,1\right]$ are given by the
definition 
\begin{equation}
\mu^{\alpha}:=\frac{1}{Z}\exp\,(-\beta\sqrt{N}y^{\alpha})
\ \ \ \
y^{\alpha}:=\frac{1}{N}\sum_{i<j}\tau_{i}^{\alpha}J_{ij}\tau_{j}^{\alpha}=\frac{1}{\sqrt{N}}H\left(\tau^{\alpha\,}\right)
\end{equation}
For example, with this notation the average of a function of $\tilde{x}\left(\sigma\right)$
is
\begin{equation}
\sum_{\sigma}\mu\left(\sigma\right)f\left(\beta\tilde{x}\left(\sigma\right)\right)=\sum_{\alpha}\,\mu^{\alpha}f\left(\beta\tilde{x}^{\alpha}\right)
\end{equation}
where we also lighted the symbols for the cavity variables 
\begin{equation}
\tilde{x}^{\alpha}:=\tilde{x}\left(\tau^{\alpha}\right),\ \ \ \tilde{y}^{\alpha}:=\tilde{y}\left(\tau^{\alpha}\right).
\end{equation}
After these manipulations the functional is as follows
\begin{equation}
A\left(\mu\right)=\log2+\log\,\sum_{\alpha}\mu^{\alpha}\cosh\left(\beta\tilde{x}^{\alpha}\right)-\log\,\sum_{\alpha}\mu^{\alpha}\exp\left(\beta\tilde{y}^{\alpha}\right)
\end{equation}
and is not (explicitly) dependent anymore on the spin orientations.

\section{Simplified ansatz}

We can introduce our simplified version of the L-RSB ansatz, which
assumes that the system at equilibrium is decomposed into a very large
(eventually infinite) number $L$ of independent REMs \cite{Franchini,Franchini2021,FranchiniSPA2,FranchiniUniversal}. Consider
the vertex set $V$, then we define the following partition 
\begin{equation}
\mathcal{V}:=\left\{ V_{1},V_{2},\,...\,,V_{\ell},\,...\,,V_{L}\right\} 
\end{equation}
that splits $V$ into $L$ disjoint subsets $V_{\ell}$, i.e., such that
\begin{equation}
V=\bigcup_{1\leq\ell\leq L}V_{\ell}
\end{equation}
summing all the subsets we get back the full system
\begin{equation}
|V|=\sum_{1\leq\ell\leq L}|V_{\ell}|=\sum_{1\leq\ell\leq L}N_{\ell}=N
\end{equation}
To match the Parisi functional exactly as is presented in \cite{Bolt} we introduce the new set sequence
\begin{equation}
Q_{\ell}=\bigcup_{\ell'\leq\ell}V_{\ell'}
\end{equation}
the sizes of these new sets are given by the sum of the parts up to $\ell$. Hereafter we set
\begin{equation}
\left|Q_{\ell}\right|=\sum_{\ell'\leq\ell}\left|V_{\ell'}\right|=q_{\ell}N.
\end{equation}
where $q_{\ell}$ is a positive non--decreasing sequence smaller than one (quantile) \cite{Franchini2021, Franchini, FranchiniSPA2, FranchiniUniversal}. 

Here is the ansatz: we assume that at equilibrium each subsystem $V_{\ell}$
behaves as an independent system governed by the Hamiltonian $H_{\ell}$.  The associated measure is a product measure
\begin{equation}
\mu\left(\sigma\right)=\prod_{\ell}\xi_{\ell}\left(\sigma_{i}:i\in V_{\ell}\right)
\end{equation}
where $\xi_{\ell}$ is the Gibbs measure associated to
the Hamiltonian $H_{\ell}$. Since the subgroups are assumed independent we can arbitrarily fix
the order in which we are going to average over them. The points of
the measure are then relabeled as follows
\begin{equation}
\mu^{\alpha}=\xi_{1}^{\alpha_{1}}\xi_{2}^{\alpha_{2}} \ ... \ \xi_{\ell}^{\alpha_{\ell}} \ ... \ \xi_{L}^{\alpha_{L}}
\end{equation}
where the sub-points $\xi_{\ell}^{\alpha_{\ell}}$ are the Gibbs measures
\begin{equation}
\xi_{\ell}^{\alpha_{\ell}}:=\frac{1}{Z_{\ell}}\exp\left(-\beta H_{\ell}\left(\tau^{\alpha_{\ell}}\right)\right)
\end{equation}
and the Hamiltonian $H_\ell$ is a that of a Random Energy Model (REM). Notice that $\xi_{\ell}^{\alpha_{\ell}}$
only acts on those variables involving spins in $V_{\ell}$ and its
support is denoted by the symbol
\begin{equation}
\Omega^{N_{\ell}}:=\{\tau^{\alpha_{\ell}}:\,1\leq\alpha_{\ell}\leq2^{N_{\ell}}\}
\end{equation}
where the state $\tau^{\alpha_{\ell}}$ is the
$\alpha_{\ell}-$th atom of the measure $\xi_\ell$ that describes the subsystem
$V_{\ell}$, 
\begin{equation}
\tau^{\alpha_{\ell}}:=\left\{ \tau_{i}^{\alpha_{\ell}}\in\Omega:\,i\in V_{\ell}\right\} .
\end{equation}

We remark that the overlap matrix of the proposed measure $\mu$ is not ultrametric. As was shown in \cite{Franchini2021, Franchini}, the proper factorization to introduce the ultrametric picture is into measures that depend not only on the spins of $V_\ell$, but also on those of previous $V_{\ell'}$ with $\ell'<\ell$ 
\begin{equation}
\mu\left(\sigma\right)=\prod_{\ell}\xi_{\ell}\left(\sigma_{i}:i\in Q_{\ell}\right)
\end{equation}
This produces the usual ultrametric structure (see Section 4 of \cite{Franchini}) of the Parisi ansatz:
\begin{equation}
\mu^{\alpha}=\xi_{1}^{\alpha_{1}}\xi_{2}^{\alpha_{1}\alpha_{2}}  \ ... \ \xi_{\ell}^{\alpha_{1}\alpha_{2} \, ... \, \alpha_{\ell}} \ ... \ \xi_{L}^{\alpha_{1}\alpha_{2} \,... \,\alpha_{L}}
\end{equation}
Remarkably, both ansatz gives the same expression for the Parisi functional. The physical reason is shown in \cite{Franchini} where it is proven that the partition function is indeed independent from the direction of the ground state, and therefore, also from the ultrametricity assumption. See Section 4 of \cite{Franchini} for the proof, and Figures 5.2 and 5.3 of \cite{Franchini} for a kernel representation of the ultrametric picture.

\section{Cavity variables}

Here we deal with the computation of the cavity variables (according
to our ansatz). The cavity field is easy, as it is natural to split 
\begin{equation}
\frac{1}{\sqrt{N}}\sum_{i}\tilde{J}_{ii}\tau_{i}^{\alpha}=\frac{1}{\sqrt{N}}\sum_{\ell}\tilde{z}_{\ell}^{\alpha_{\ell}}\sqrt{\left|V_{\ell}\right|}
\end{equation}
into independent variables that are functions of the $V_{\ell}$ spins only
\begin{equation}
\tilde{z}_{\ell}^{\alpha_{\ell}}\sqrt{\left|V_{\ell}\right|}:=\sum_{i\in V_{\ell}}\tilde{J}_{ii}\tau_{i}^{\alpha_{\ell}}
\end{equation}
The correction is more subtle, but we can still decompose it into an adapted sequence. Call 
\begin{equation}
W:=\left\{ \left(i,j\right):\,1\leq i,j\leq N\right\} 
\end{equation}
the edges set associated to $V$, then $\mathcal{V}$ induces a partition of $W$ into subsets $W_{\ell}$ \cite{Franchini2021, Franchini} such that each $W_{\ell}$ contains all edges with both ends in $Q_{\ell}$ minus those with both ends in $Q_{\ell-1}$: see Figure 1 of \cite{Franchini2021} or Figure 4.1 of \cite{Franchini}. By introducing the following variable \cite{Franchini2021, Franchini}
\begin{equation}
\tilde{g}_{\ell}^{\alpha_{1}...\alpha_{\ell}}\sqrt{\left|W_{\ell}\right|}:=\sum_{i\in V_{\ell}}\sum_{j\in V_{\ell}}\tau_{i}^{\alpha_{\ell}}\tilde{J}_{ij}\tau_{j}^{\alpha_{\ell}}+\sum_{i\in V_{\ell}}\sum_{\ell'<\ell}\sum_{j\in V_{\ell'}}\tau_{i}^{\alpha_{\ell}}\tilde{J}_{ij}\tau_{j}^{\alpha_{\ell'}}+\sum_{j\in V_{\ell}}\sum_{\ell'<\ell}\sum_{i\in V_{\ell'}}\tau_{i}^{\alpha_{\ell'}}\tilde{J}_{ij}\tau_{j}^{\alpha_{\ell}}
\end{equation}
we can write the decomposition
\begin{equation}
\frac{1}{N}\sum_{i<j}\tau_{i}^{\alpha}\tilde{J}_{ij}\tau_{j}^{\alpha}=\frac{1}{\sqrt{2}}\sum_{\ell}\sum_{\left(i,j\right)\in W_{\ell}}\tau_{i}^{\alpha}\tilde{J}_{ij}\tau_{j}^{\alpha}=\frac{1}{\sqrt{2}N}\sum_{\ell}\tilde{g}_{\ell}^{\alpha_{1}...\alpha_{\ell}}\sqrt{\left|W_{\ell}\right|}
\end{equation}
where $1/\sqrt{2}$ comes from removing the $i<j$ constraint under
the assumption that $\tilde{J}$ is asymmetric almost surely. 
Notice that, since $\tilde{J}$ is independent from $J$, we expect that the
kernel of $\mu$ does not diagonalize the Hamiltonian $\tilde{H}$, and therefore we can consider the sub-states of $V_{\ell}$ as if they where independent from each other. Then the following holds in distribution:
\begin{equation}
\tilde{g}_{\ell}^{\alpha_{1}...\alpha_{\ell}}\stackrel{d}{=}\tilde{g}_{\ell}^{\alpha_{\ell}},
\end{equation}
with both $\tilde{z}_{\ell}^{\alpha_{\ell}}$ and $\tilde{g}_{\ell}^{\alpha_{\ell}}$
normally distributed with unitary variance for all $\ell$.
Now compute
\begin{equation}
\left|V_{\ell}\right|=\left|Q_{\ell}\right|-\left|Q_{\ell-1}\right|=\left(q_{\ell}-q_{\ell-1}\right)N,
\ \ \ 
\left|W_{\ell}\right|=\left|Q_{\ell}\right|^{2}-\left|Q_{\ell-1}\right|^{2}=\left(q_{\ell}^{2}-q_{\ell-1}^{2}\right)N^{2}.
\end{equation}
Then, the cavity variables are given by
\begin{equation}
\tilde{x}^{\alpha}\stackrel{d}{=}\sum_{\ell}\tilde{z}_{\ell}^{\alpha_{\ell}}\sqrt{q_{\ell}-q_{\ell-1}},
\ \ \ \
\tilde{y}^{\alpha}\stackrel{d}{=}\frac{1}{\sqrt{2}}\sum_{\ell}\tilde{g}_{\ell}^{\alpha_{\ell}}\sqrt{q_{\ell}^{2}-q_{\ell-1}^{2}}.
\end{equation}
Putting together the functional becomes
\begin{multline}
A\left(\mu\right)\stackrel{d}{=}\log2\,+\log\,\sum_{\alpha_{1}}\,\xi_{1}^{\alpha_{1}}...\,\sum_{\alpha_{L}}\,\xi_{L}^{\alpha_{L}}\cosh\left(\beta\sum_{\ell}\tilde{z}_{\ell}^{\alpha_{\ell}}\sqrt{q_{\ell}-q_{\ell-1}}\right)\\
-\log\,\sum_{\alpha_{1}}\,\xi_{1}^{\alpha_{1}}...\,\sum_{\alpha_{L}}\,\xi_{L}^{\alpha_{L}}\exp\left(\frac{\beta}{\sqrt{2}}\sum_{\ell}\tilde{g}_{\ell}^{\alpha_{\ell}}\sqrt{q_{\ell}^{2}-q_{\ell-1}^{2}}\right)\label{eq:ss}
\end{multline}
It only remains to discuss the properties of the sub-points $\xi_{\ell}^{\alpha_{\ell}}$
under the assumption that these are approximated by REM Gibbs measures.

\section{REM and the Parisi formula}

If the $H_\ell$ are Gaussian random energies (a Random Energy Model, REM) \cite{Kurkova1,Arous-Kupsov}, then the full probability measure $\mu$ is obtained from a product of probability measures of the kind 
\begin{equation}
\xi_{\ell}^{\alpha_{\ell}}=\frac{\eta_{\ell}^{\alpha_{\ell}}}{\sum_{\gamma{}_{\ell}}\eta_{\ell}^{\gamma{}_{\ell}}}
\end{equation}
where $\eta_{\ell}^{\alpha_{\ell}}$ are points of a Poisson
Point Process (PPP) with
rate parameter $\lambda_{\ell}$. Then, by the special average properties of the PPP \cite{ASS,Bolt,PaK},
also known as Little Theorem in \cite{Mezard}, it is possible to prove the following average formula:
\begin{equation}
\sum_{1\leq\alpha_{\ell}\leq2^{N_\ell}}\xi_{\ell}^{\alpha_{\ell}}f\left(\tilde{z}_{\ell}^{\alpha_{\ell}}\right)=K_{\ell}\left(\frac{1}{2^{N_\ell}}\sum_{\alpha_{\ell}}f\left(\tilde{z}_{\ell}^{\alpha_{\ell}}\right)^{\lambda_{\ell}}\right)^{\frac{1}{\lambda_{\ell}}}
\end{equation}
for some constant $K_{\ell}$. This allows to compute the main contribution
\begin{equation}
\sum_{\alpha_{1}}\,\xi_{1}^{\alpha_{1}}...\,\sum_{\alpha_{L}}\,\xi_{L}^{\alpha_{L}}\cosh\left(\beta\sum_{\ell}\tilde{z}_{\ell}^{\alpha_{\ell}}\sqrt{q_{\ell}-q_{\ell-1}}\right)=Y_{0}\exp\left(\sum_{\ell}\log K_{\ell}\right)
\end{equation}
by applying the following chain of equations
\begin{equation}
Y_{\ell-1}=K_{\ell} {\left( \frac{1}{2^{N_\ell}}\sum_{\alpha_{\ell}}Y_{\ell}^{\lambda_{\ell}}\right)^{\frac{1}{\lambda_{\ell}}}}
\end{equation}
from the initial condition at step $\ell=L$,
\begin{equation}
Y_{L+1}=\cosh\left(\beta\sum_{\ell}\tilde{z}_{\ell}^{\alpha_{\ell}}\sqrt{q_{\ell}-q_{\ell-1}}\right),
\end{equation}
down to $\ell=0$. Then we compute the correction therm in the same
way, finding 
\begin{multline}
\sum_{\alpha_{1}}\,\xi_{1}^{\alpha_{1}}...\,\sum_{\alpha_{L}}\,\xi_{L}^{\alpha_{L}}\exp\left(\frac{\beta}{\sqrt{2}}\sum_{\ell}\tilde{g}_{\ell}^{\alpha_{\ell}}\sqrt{q_{\ell}^{2}-q_{\ell-1}^{2}}\right)\\
=\exp\left(\frac{\beta^{2}}{4}\sum_{\ell}\lambda_{\ell}\left(q_{\ell}^{2}-q_{\ell-1}^{2}\right)+\sum_{\ell}\log K_{\ell}\right)\label{eq:vv}
\end{multline}
Putting together the contributions depending from $K_{\ell}$ cancel out and we finally obtain the Parisi functional as formulated in \cite{ASS,Bolt,PaK}:
\begin{equation}
A\left(\mu\right)\stackrel{d}{=}\log2+\log Y_{0}-\frac{\beta^{2}}{4}\sum_{\ell}\lambda_{\ell}\left(q_{\ell}^{2}-q_{\ell-1}^{2}\right).
\end{equation}

Also, see the Section 4.2.1 of \cite{Bardella2024} for how to obtain the RSB theory as special (stationary) case of the kernel picture \cite{Franchini,Franchini2021,FranchiniSPA2,FranchiniUniversal} within the larger context of Lattice Field Theories \cite {iScience2024}.

\section{Acknowledgments}

We would like to thank Giorgio Parisi and Riccardo Balzan (Sapienza Univerista
di Roma) and Pan Liming (UESTC) for interesting discussions and suggestions.
This project has received funding from the European Research Council
(ERC) under the European Unions Horizon 2020 research
and innovation programme (grant agreement No 694925).

\end{document}